\documentclass[11pt]{article}
\pdfoutput=1
\usepackage{amsmath}
\usepackage{amssymb}
\usepackage{graphicx,bbm,mathrsfs}
\usepackage{nicefrac}
\usepackage{slashed}
\usepackage{bbm}
\usepackage{geometry}
\usepackage{stackrel}
\usepackage{setspace}
\usepackage{relsize}
\usepackage{pifont}% http://ctan.org/pkg/pifont
\usepackage{mathtools}
\usepackage{enumitem}
\usepackage{dcolumn}   % needed for some tables
\usepackage{bm}        % for math
\usepackage{graphicx,mathrsfs}
\usepackage{nicefrac}
\usepackage{multirow}
\usepackage{color}
\usepackage{mathtools}
\usepackage{makecell}
\usepackage{environ} 
\usepackage{lipsum} 
\usepackage{wasysym}
\usepackage{hhline,colortbl}  
\usepackage{tikz}
\usepackage{graphicx}
\usepackage{tensor}
\usepackage{dsfont}
\usepackage{simpler-wick}
\usepackage{hyperref}

\newcommand{\MM}{\mathcal{M}}
\newcommand{\ii}{\mathrm{i}}
\newcommand{\nn}{\mathbf{n}}

\newcommand{\dd}{\mathrm{d}}
\newcommand{\Beta}{\mathrm{B}}
\newcommand{\be}{\begin{equation}}
\newcommand{\ee}{\end{equation}}
\newcommand{\bea}{\begin{eqnarray}}
\newcommand{\eea}{\end{eqnarray}}
\newcommand{\D}{\mathrm{D}}

\newcommand{\ta}{\Tilde{a}}
\newcommand{\tf}{\Tilde{f}}
\newcommand{\tv}{\Tilde{v}}

\begin{document}

\vspace*{4mm}

\thispagestyle{empty}

\begin{center}

%  {\LARGE
% \sc
\begin{minipage}{20cm}
\begin{center}
\hspace{-5cm }
\huge
\sc
\\
\hspace{-5cm } 
Vortex Fractional Fermion Number\\
\hspace{-5cm } through Heat Kernel methods and Edge States
\end{center}
\end{minipage}
\\[30mm]

\renewcommand{\thefootnote}{\fnsymbol{footnote}}

{\large  
Sylvain~Fichet$^{\, a}$ \footnote{sylvain.fichet@gmail.com}\,, 
Rodrigo~Fresneda$^{\, b}$ \footnote{rodrigo.fresneda@ufabc.edu.br}\,,
Lucas~de~Souza$^{\, b}$ \footnote{souza.l@ufabc.edu.br}\,, 
Dmitri Vassilevich$^{\, b}$ \footnote{dvassil@gmail.com}
}\\[12mm]
\end{center} 
\noindent

\indent \; ${}^a\!$ 
\textit{CCNH, Universidade Federal do ABC,} \textit{Santo Andr\'e, 09210-580 SP, Brazil}

\indent \; ${}^b\!$ 
\textit{CMCC, Universidade Federal do ABC,} \textit{Santo Andr\'e, 09210-580 SP, Brazil}
\\

\addtocounter{footnote}{-4}

\vspace*{8mm}
 
\begin{center}
{  \bf  Abstract }
\end{center}
\begin{minipage}{15cm}
\setstretch{0.95}
% \small

Computing the vacuum expectation of fermion number operator on a soliton background is often challenging. A recent proposal in \cite{Fresneda:2023wub} simplifies this task by considering the soliton in a bounded region and relating the $\eta$ invariant, and thus the fermion number, to a specific heat kernel coefficient and to contributions from the edge states. We test this method in a system of charged fermions living on an Abrikosov–Nielsen–Olesen (ANO) vortex background. 
% We show that the resulting $\eta$ invariant is  independent of boundary conditions, confirming the validity of the method. 
We show that the resulting $\eta$ invariant does not depend on boundary conditions (within a certain class), thereby supporting the 
validity of the method. Our analysis reveals a  nontrivial feature for the fermionic spectrum in the vortex-induced Higgs phase. 
As a by-product, we also find that for a vortex living on a disk,  the edge states carry fractional charge.  

    \vspace{0.5cm}
\end{minipage}

\newpage
\setcounter{tocdepth}{2}

\tableofcontents  

\vspace{1cm}
\hrule
\vspace{1cm}

\date{\empty}

\section{Introduction}
It was discovered a long time ago by Jackiw and Rebbi \cite{Jackiw:1975fn} that the vacuum expectation value of the fermion number operator on the background of a soliton may be non-integer. Soon afterwards, many other examples of fermion number fractionization were discovered, see e.g. \cite{Goldstone:1981kk,Jackiw:1981wc,Paranjape:1983dy} and the review \cite{Niemi:1984vz}. An important relation between quantum anomalies and fermion fractionization was established in \cite{Niemi:1983rq}.

The Abrikosov--Nielsen--Olesen (ANO) vortex \cite{Abrikosov:1956sx,Nielsen:1973cs} is probably the best known solitonic configuration in $2+1$ dimensions. The ANO vortex is particularly important for the theory of type II superconductivity, for some recent development see \cite{Kim:2025ien}. Quantum properties of the vortex have been a subject of intensive studies over many years. The one-loop mass shift of the vortex was calculated in \cite{Vassilevich:2003xk,Rebhan:2003bu} in the supersymmetric case, while the bosonic and fermionic contributions separately were considered in \cite{Bordag:2003at,AlonsoIzquierdo:2004ru,Graham:2004jb,Alonso-Izquierdo:2016bqf}, see \cite{Graham:2022adn} for some recent results. The vacuum expectation value of fermion number operator depends crucially on the coupling of fermions to the bosonic background. With the choice of coupling made in \cite{Chamon:2007hx} the fermion number $\mathcal{N}$ was calculated to be $\pm n/2$ with $n$ being the topological charge of the vortex. On the other hand, for the couplings in the Jackiw--Rossi model \cite{Jackiw:1981ee} the fermion number appears to be non-topological \cite{Almeida:2021lks}.

% The main method for calculating the vacuum expectation value of the fermion number operator is the derivative expansion. 
The main method for calculating the vacuum expectation value of the fermion number operator is the derivative expansion. This procedure can be nicely organized as a resummation of the heat kernel expansion \cite{Alonso-Izquierdo:2019tms}. Even then, the calculations may be rather complicated and one can encounter convergence issues. 

There is a different method \cite{Fresneda:2023wub} which avoids the convergence problems and simplifies the combinatorics. It is based on the observation that the variation of $\mathcal{N}$ under local variations of background bosonic fields is given by a very simple explicit formula through a heat kernel coefficient. Although locality is a very mild restriction, it excludes variations which change the asymptotic behavior of the fields. As we will see below, this makes the variations formula
unsuitable on $\mathbb{R}^2$. Thus, it is natural to consider the vortex on a disk of large but finite radius $R$, compute $\mathcal{N}$ for this system, and take the limit $R\to\infty$. However, by introducing a boundary one introduces a near-boundary fermion density which may give a non-vanishing contribution to $\mathcal{N}$ even in the limit $R\to\infty$. This contribution has to be subtracted from the final answer.  It is natural to assume (though hard to prove rigorously) that the boundary fermion density is defined at large $R$ by the edge states. Since the boundary is compact (a circle) the variational formula is again applicable. In other words, this scheme allows to express a complicated quantity through a difference of two much simpler quantities.

In this work, we claim to achieve the following goals. 
\begin{enumerate}
\item We check the method proposed in \cite{Fresneda:2023wub}. The method  passes several non-trivial consistency checks for different choices of boundary conditions.
\item Our results  help to reveal the quantum structure on the ANO vortex with massive charged fermions. We also make predictions for the zero-mode structure.
\item As a by-product, we  find the spectrum of edge states. It appears that these states are fractionally charged, which may have some applications to the theory of Fractional Quantum Hall Effect \cite{MacDonald:1990zz,Lopez:1998ih}, for example.
\end{enumerate}

The paper is structured as follows. In section \ref{sec:set} we introduce the setup of fermions on  the vortex  and present the general calculation method of ${\cal N}$. We then apply the method to the vortex in section \ref{sec:calc} and discuss the results in section \ref{sec:dis}.

\section{Setup and Method}\label{sec:set}

\subsection{Fermions and the ANO Vortex}

\label{se:setup}

In this work, we consider the following Lagrangian density for fermions,
\begin{equation}
\mathcal{L}=\ii \bar\psi \gamma^\mu D_\mu \psi + \ii \bar \chi \gamma^\mu \partial_\mu \chi - \ii\sqrt{2}e (\bar\psi \chi \phi -\bar\chi \psi \phi^*) - m(\bar\psi\psi + \bar\chi\chi)\,, \label{ANOL}
\end{equation}
where $\psi$ and $\chi$ are complex spinors, and $\phi$ is a complex scalar field. The fields $\psi$ and $\phi$ have electric charge $+e$, and the covariant derivative is $D_\mu=\partial_\mu -\ii e A_\mu$. The field $\chi$ is uncharged. Except for the mass term, this Lagrangian corresponds to the fermionic part of the $N=2$ supersymmetric Higgs model in $2+1$ dimensions. The $2 + 1$ dimensional metric is taken to be $(1, -g^{ij} )$, where $g^{ij}$ is a flat positive metric on $\mathbb{R}^2$.

As a classical background we consider the ANO vortex
\begin{equation}
\phi=f(r)e^{-\ii n \theta},\qquad eA_j=\epsilon_{jk}\frac{x^k}{r^2} (a(r)-n), \label{ANOs}
\end{equation} 
where $\epsilon_{jk}$ is the Levi-Civita tensor. In polar coordinates on $\mathbb{R}^2$ we consider the clockwise orientation, i.e., $\epsilon_{r\theta}=-r$. The functions $f$ and $a$ satisfy the boundary conditions
\begin{equation}
a(0)=n,\quad a(\infty)=0,\quad f(0)=0,\quad f(\infty)=v,\label{ANObc}
\end{equation}
and the equations
\begin{equation}
\frac 1r \frac {\rm d}{{\rm d}r} a(r)=e^2(f(r)^2-v^2), \qquad
r \frac {\rm d}{{\rm d}r} \ln f(r) =a(r).
\end{equation}
Here $v$ is a positive constant corresponding to a minimum of the Higgs potential.

It is important for our purposes that both $a(r)$ and $f(r)$ approach their asymptotic values at $r\to\infty$ prescribed in (\ref{ANObc}) exponentially fast, while $|D_j\phi|$ and $F_{ij}F^{ij}$ decay exponentially fast in this limit. It is easy to check that
\begin{equation}
\frac e{4\pi}\int \mathrm{d}^2x\, F_{ij}\epsilon^{ij}=n, \label{ANOn}
\end{equation}
where $F_{i j} = \partial_i A_j - \partial_j A_i$, so that $n\in\mathbb{Z}$ is a topological charge of the ANO vortex.

Let us introduce the 4-spinor
\begin{equation}
\Psi =\begin{pmatrix}
\psi \\ \chi
\end{pmatrix}.
\end{equation}
The Dirac Hamiltonian corresponding to (\ref{ANOL}) and acting on $\Psi$ reads
\begin{equation}
H = - \begin{pmatrix}
\ii\alpha^j (\partial_j - \ii eA_j) - \beta m& -\ii e\sqrt{2} \phi \beta \\
\ii e\sqrt{2} \phi^{*}\beta & \ii\alpha^j \partial_j - \beta m
\end{pmatrix}, \label{ANOH2}
\end{equation}
where $\alpha^j=\gamma^0\gamma^j$, $\beta=\gamma^0$. These matrices satisfy the algebraic relations
\begin{equation}
\mathrm{tr}\, \left(\alpha^k\alpha^j\beta \right)=-2\ii \epsilon^{kj}, \qquad \alpha^j\alpha^k + \alpha^k\alpha^j = 2g^{jk}, \label{tr}
\end{equation}
Besides, $\beta\alpha^j+\alpha^j\beta=0$ and $\beta^2=1$.

\subsection{Boundary Conditions}

We will work on the infinite space $\mathbb{R}^2$ as well as on the disk $\D_R$ of large radius $R$. On the boundary of $\D_R$ we will impose bag boundary conditions
\begin{equation}
\Pi_{\varepsilon}\Psi\vert_{\partial\D_{R}}=0, \label{bc}
\end{equation}
with
\be
 \Pi_{\varepsilon}=\frac{1}{2}(1- X_{\varepsilon})\,, \quad X_{\varepsilon}= \left(\begin{array}{cc}
    \ii \varepsilon  \beta \alpha^{\mathbf{n}}  & 0 \\
    0        & \ii \varepsilon  \beta \alpha^{\mathbf{n}}
\end{array} \right), 
 \label{BC}\ee
where  $\varepsilon = \pm 1$ corresponds to two inequivalent choices of boundary conditions. 
Here $\alpha^\nn\equiv e^\nn_j\alpha_j$, where $e^\nn$ is an inward pointing unit normal to the boundary. Additionally, let $e^{\|}$ be a unit vector tangential to the boundary, so that, with $\alpha^{\|}:=e_j^{\|} \alpha^j$.  Near the boundary, $x^\nn=R-r$ and $x^\parallel = R \theta$.
For these conditions, the normal components of the fermion currents, $\psi^\dag \alpha^\mathbf{n} \psi$ and $\chi^\dag \alpha^\mathbf{n} \chi$, vanish at all points of the boundary, so that the Dirac Hamiltonian is self-adjoint. 

We will perform the calculations for both boundary conditions, i.e. both values of $\varepsilon$. Independence of the final result from the choice of $\varepsilon$ serves as  a consistency check of the formalism.
\subsection{Computing the Fermion Number}
\label{se:method}

The expectation value of the fermion number for a theory with Dirac Hamiltonian $H$ reads \cite{Niemi:1983rq}
\begin{equation}
\mathcal{N}(H)=-\tfrac 12 \eta(0,H)\,, \label{Neta}
\end{equation}
where the spectral $\eta$ function of $H$ is defined as
\begin{equation}
\eta(s,H)=\sum_{\lambda>0}\lambda^{-s} - \sum_{\lambda<0}(-\lambda)^{-s}, \label{etas}
\end{equation}
with $\lambda$ being the eigenvalues of $H$. Here $s$ is a complex spectral parameter.

If under a variation of the Hamiltonian one zero mode appears or disappears, then $\eta(0,H)$ jumps by $\pm 1$. If a mode crosses the origin, $\eta(0,H)$ jumps by $\pm 2$. If, on the contrary, the signs of all modes remain unchanged under the variation $H\to H+\delta H$, the $\eta$ function changes smoothly and the variation can be expressed as
\cite{AlvarezGaume:1984nf,Atiyah:1980jh,Gilkey:1984} \footnote{For a derivation of this formula, see e.g. (A.7)-(A.10) in \cite{AlvarezGaume:1984nf} or (2.8) - (2-12) in \cite{Kurkov:2018pjw}.}
\begin{equation}
\delta\eta(0,H)=-\frac 2{\sqrt{\pi}} a_{d-1}(\delta H,H^2), \label{vareta}
\end{equation}
by means of a coefficient in the heat kernel expansion for $H^2$,
\begin{equation}
\mathrm{Tr}\, \left( q e^{-tH^2}\right) \simeq \sum_{k=0}^\infty t^{\frac{k-d}{2}} a_k(q,H^2)  \,.\label{asymhk}
\end{equation}
Here $d$ is the dimensionality of the manifold, and $q$ is a matrix-valued function. We have in particular,
% the identities \cite{Vassilevich:2003xt},
\begin{eqnarray}
&&a_0(q,H^2)=\frac{1}{(4\pi)^{\frac{d}{2}}}\int_\MM \dd^dx\,\sqrt{g}\,\mathrm{tr}\, (q),\label{a0}\\
&&a_1(q,H^2)=\frac 1{4(4\pi)^{\frac{d-1}{2}}} \int_{\partial \MM} \dd^{d-1}x\, \sqrt{h}\, \mathrm{tr}\, ( q X )\,,\label{a1}
\end{eqnarray}
% \new{see Eqs. (5.29) and (5.30) in \cite{Vassilevich:2003xt}. $X_{\varepsilon}$ depends on the boundary conditions and is given in the formula (\ref{BC}) above.}
where $\dd^{d-1}x\sqrt{h}$ is the induced integration measure on the boundary $\partial M$ and $X$ depends on the boundary conditions, see Eqs. (5.29) and (5.30) in \cite{Vassilevich:2003xt}.   In the present work, $X\equiv X_\varepsilon$, which is defined in \eqref{BC}. 
% Here $X_{\varepsilon}$ depends on the boundary conditions as defined in \eqref{BC}.
% %
% In the case of bag boundary conditions (\ref{bc}), each choice of the parameter $\varepsilon=\pm 1$ specifies a distinct boundary condition.
%

It is important to note that the variation $\delta H$ in the above formulae has to be local. That is, it has to vanish exponentially fast outside of a compact domain. This requirement does not impose any restrictions if the spectral problem is considered on a compact manifold with or without boundaries. However, on $\mathbb{R}^2$ it means that one cannot vary the asymptotics of the background fields. 

The vacuum fermion number can be represented as an integral of a local density $\rho(x)=\langle \psi^\dag (x)\psi (x) +\chi^\dag (x)\chi(x)\rangle$,
\begin{equation}
\mathcal{N}(H)=\int_{\mathbb{R}^2} \dd^2x\, \rho(x)\,. \label{Nrho}
\end{equation}
Let us restrict the integration in (\ref{Nrho}) to a disk $\D_R$ of radius $R$ centered at the origin and define
\begin{equation}
\mathcal{N}_R(H)=\int_{\D_R} \dd^2x\, \rho(x)\,. \label{NDRrho}
\end{equation}
Let us consider a theory described by a Hamiltonian $H_{\D_R}$ which has the same symbol as $H$ but acts on the spinors satisfying some boundary conditions on $S^1_R=\partial \D_R$. Since we consider only localized solitonic backgrounds, we expect that for sufficiently large $R$ the fermion density corresponding to $H_{\D_R}$ coincides with $\rho(x)$ for $H$ everywhere except for a vicinity of the boundary. It is natural to assume that the near-boundary contribution is given by the fermion number of the effective boundary Hamiltonian $H_{\mathrm{b}}$  acting on edge states, i.e.
\begin{equation}
\mathcal{N}_R(H) = \mathcal{N}(H_{\D_R})-\mathcal{N}(H_{\mathrm{b}}), \qquad R\to\infty.\label{NNN}
\end{equation} 
By definition, edge states are the eigenmodes of $H_{\mathrm{D}_R}$ which decay as $e^{-\mu x^n}$ with $\Re \mu>0$ and $x^{\mathbf{n}}$ measures the distance from the boundary. If the decay rate of all edge states is bounded from below by a positive constant $\mu_0$,  $\Re \mu>\mu_0>0$, one can guarantee that edge states do not contribute to the fermion density for $x^{\mathbf{n}} \gg \mu_0^{-1}$. These arguments justify (\ref{NNN}) though do not provide a rigorous proof. The existence of $\mu_0$ depends on boundary conditions. For the conditions used in this paper such limit indeed exists except for some critical values of the parameters, see below Eq. (\ref{detEq}). Eq. (\ref{NNN}) has been checked
and confirmed in \cite{Fresneda:2023wub} for the example of a planar fermion in an external magnetic field. 
This calculation method is summarized in Fig.\,\ref{MethodFigure}.

In practice, we will analyze \emph{smooth} parts\footnote{\begin{minipage}[t]{\columnwidth}Formally, this means transitioning from the $\eta$ function to the so-called exponentiated $\eta$ invariant, which is a smooth function of background fields, see \cite{Fresneda:2023wub}.\end{minipage}} of the variations of $\mathcal{N}$ (given by (\ref{Neta}) and (\ref{vareta})) caused by variations of $A_j$, $\phi$ and $m$. This is given by  the equation
\begin{equation}
\delta \mathcal{N}_R(H) = \delta \mathcal{N}(H_{\D_R})-\delta \mathcal{N}(H_{\mathrm{b}}), \qquad R\to\infty \label{delNNN}
\end{equation}
following from (\ref{NNN}), and its counterpart for variations of the $\eta$ function
\begin{equation}
\delta \eta(0,H)_R = \delta \eta(0,H_{\D_R})-\delta \eta(0,H_{\mathrm{b}}), \qquad R\to\infty.\label{deleee}
\end{equation}
This trick allows us to circumvent the necessity of analyzing zero modes. As we will see below, the equations (\ref{delNNN}) and (\ref{deleee}) can be solved uniquely for our model.

Note that the variational formula ({\ref{vareta}) can be applied directly to $\eta(0,H)$ on $\mathbb{R}^2$. Since $a_1$  does not contain any bulk terms (see  (\ref{a1})),  we can only conclude that $\eta(0,H)$ is stable against any local variations of the background field, i.e., $\eta(0,H)$ is a topological invariant.

\begin{figure}[t]
    \centering
\includegraphics[scale=0.5]{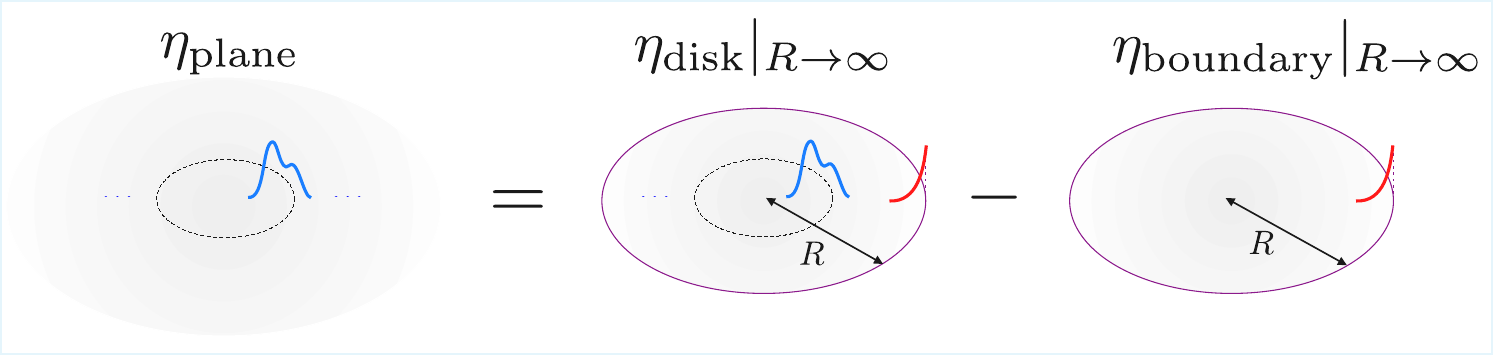}
\caption{ Calculation method of the $\eta$ or $\cal N$ invariants on a solitonic configuration in ${\mathbb R}^2$ {\it (left)}. We put the configuration on a disk of radius $R$ {\it (middle)},  subtract the contribution from the edge states {\it (right)} and take $R\to\infty$.  
%A pictorial representation of \new{the integrated version of} Eq.\ (\ref{deleee}). 
% The \old{bubble} on the left hand side is the $\eta(0,H)$ on $\mathbb{R}^2$. The term in the middle is the disk contribution, while that of right is the contribution edge states.
}
    \label{MethodFigure}
\end{figure}

\section{Calculating the   $\eta$ Invariant on a  Vortex  }\label{sec:calc}

We apply the method of section \ref{se:method} to the fermions living on the vortex background described in section \ref{se:setup}.

\subsection{The $\eta$ Invariant on the Disk}\label{sec:disk}

We compute the $\eta$ invariant on the disk using (\ref{vareta}) and (\ref{a1}). 

Let us consider small variations of the gauge field, mass, and the Higgs field. The resulting variation of the Dirac Hamiltonian reads
\begin{equation}
    \delta H_{\D_R} = e\delta A_j \begin{pmatrix}
        \alpha^j & 0\\
        0 & 0
    \end{pmatrix} + \delta m \begin{pmatrix}
        \beta & 0\\
        0 & \beta
    \end{pmatrix} + \begin{pmatrix}
        0 & \ii e\sqrt{2}\beta  \delta(\phi)\\
      -\ii e\sqrt{2}\beta \delta(\phi^{*})   & 0
    \end{pmatrix}.
\end{equation}
By computing the traces, we easily obtain
\begin{equation}
\mathrm{tr}( \delta H_{\D_R}  X )=-2\varepsilon\,e\, \delta A_j \epsilon^{\mathbf{n} j}\,,
\end{equation}
where $\epsilon^{\nn j} = e^\nn_i \epsilon^{i j}$. Thus, by (\ref{vareta}) and (\ref{a1}), one has
\begin{equation}
\delta\eta(0,H_{\D_R})=\varepsilon\frac{e}{2 \pi}\int_{S^1_R} \dd \theta\, \sqrt{h}\, \delta A_j \epsilon^{\mathbf{n} j} \,,\label{delHDR}
\end{equation}
where $\sqrt{h} \dd \theta =R \dd \theta$ is the induced integration measure. Our choice of orientation corresponds to $ e^{\parallel}_j \epsilon^{\nn j} =+1$.

\subsection{Contribution from the Edge States}\label{sec:edge}

We compute the contribution from the edge states at the boundary of the disk, i.e. the right term in Fig.\,\ref{MethodFigure}.

To find the edge states, we suppose that $R$ is large and restrict the Hamiltonian to a vicinity of the boundary $S_R^1$, where we consider the coordinate $x^{\parallel} = R \theta$. In this vicinity, we perform a gauge transformation
\bea
\phi & \rightarrow &\phi^{[\alpha]}= e^{\ii \alpha(x)  }\phi \nonumber \\
\psi & \rightarrow &\psi^{[\alpha]}= e^{\ii \alpha(x)  }\psi \nonumber \\
\chi & \rightarrow & \chi \nonumber \\
A_{j} & \rightarrow & A_j^{[\alpha]}= A_{j} + \frac{1}{e} \partial_{j}\alpha(x) \label{gautr}
\eea
with $\alpha(x) =  n \theta$. This transformation removes oscillations of the phase of field $\phi$ and the $1/r$ tail in $A_j$. Note that (\ref{gautr}) cannot be extended continuously to the whole disk $\D_R$.

The background bosonic fields become slowly varying near the boundary and thus can be replaced by their values at $r=R$,
\begin{equation}
eA_{\parallel}^{[\alpha]}=\frac{a(R)}{R}\,,\qquad \phi^{[\alpha]}=f(R)\,.\label{valuesatR}
\end{equation}
One can see that $A_\nn=0$ (and also $A_{\nn}^{[\alpha]}=0$) for the ANO vortex. 

The Dirac Hamiltonian in the near-boundary problem takes the form
\begin{equation}
\hat{H} = - \begin{pmatrix}
\ii\alpha^{\mathbf{n}}  \partial_{\mathbf{n}}  + \alpha^{\parallel} (\kappa + \ta)  - \beta m&   - \ii \tf \beta  \\
   \ii \tf \beta & \ii\alpha^{\mathbf{n}} \partial_{\mathbf{n}} + \alpha^{\parallel} \kappa  - \beta m 
\end{pmatrix}, 
\end{equation}
where 
\begin{equation}
    \ta = e A_{\parallel}^{[\alpha]}(R)\,, \quad \tf =e \sqrt{2} f(R)\,.\label{PHIbar}
\end{equation}
For the ANO background, $\ta$ and $\tf$ are given by (\ref{valuesatR}). We allow for small (constant) deviations from these values. We do not allow fluctuations of $A_\nn$ which can be considered as a gauge condition on variations of background fields.
Since $\hat{H}$ commutes with $\partial_\parallel$, we have replaced the tangential derivative with its eigenvalue, 
$\ii\partial_\parallel \Psi=\kappa\Psi$. Later on, we will also need a notation for the asymptotic value of $\tf$ at $R\to\infty$,
\begin{equation}
\tv = e \sqrt{2} v\,.
\end{equation}

The eigenvalue problem
\be
\hat{H} \Psi = \lambda \Psi\,, \label{AuxProblemH} 
\ee
can be rewritten as
\begin{equation}
\big(\partial_{\mathbf{n}} + Q\big)\Psi= 0\,,  \label{AuxProblemQ}
\end{equation}
with 
\begin{equation}
  Q =\begin{pmatrix}
         -\ii\alpha^{\mathbf{n}}( \alpha^{\parallel} (\kappa + \ta) - \beta m + \lambda) & -\ii \alpha^{\Vert} \tf \\ \ii \alpha^{\Vert} \tf&  -\ii\alpha^{\mathbf{n}}( \alpha^{\parallel} \kappa  - \beta m + \lambda) 
    \end{pmatrix}.  \label{Qmatrix}  
\end{equation}

Edge states are the solutions of (\ref{AuxProblemQ}) which satisfy boundary conditions and decay exponentially as functions of $x^\nn$. Let us fix the following representation of the Dirac matrices,
\begin{equation}
\beta = \begin{pmatrix}
1 & 0 \\
0 & -1
\end{pmatrix}, \quad \alpha^\mathbf{n} = \begin{pmatrix}
0 & 1 \\
1 & 0
\end{pmatrix}, \quad \alpha^\parallel = \begin{pmatrix}
0 &\ii\\
-\ii & 0
\end{pmatrix}. 
\label{DiracRep} 
\end{equation}
The matrix $Q$ has four eigenvalues $\pm \mu_\pm$, 
\begin{equation}
    \mu_{\pm} \approx \sqrt{(m\pm\tf)^2+\kappa^2 -\lambda^2}\,,
\label{decayRate}\end{equation}
where we take the positive value of the square root. If the expression under the square root is negative, the corresponding eigenvalue is imaginary and thus does not give an edge state. We have neglected the corrections to (\ref{decayRate}) which vanish exponentially fast at $R \rightarrow \infty$. The general solution of (\ref{AuxProblemQ}) reads 
\begin{equation}
   \Psi(x^{\mathbf{n}}) = C_{1} e^{\mu_{-}x^{\mathbf{n}}}u_{-} +  C_{2} e^{-\mu_{-}x^{\mathbf{n}}}w_{-}+  C_{3} e^{\mu_{+}x^{\mathbf{n}}}u_{+}  +  C_{4} e^{-\mu_{+}x^{\mathbf{n}}}w_{+}\,, \label{AuxProbQSolution}
\end{equation} 
where $u_{\pm}$ and $w_{\pm}$ are eigenvectors of $Q$ corresponding to its negative and positive eigenvalues, respectively, and $C_{\ell}$ are constants. The modes decaying at $x^\nn\to\infty$ correspond to $w_-$ and $w_+$,
\be
w_{-} \approx \begin{pmatrix}
    \frac{\kappa -\sqrt{(m-\tf)^2+\kappa^2 -\lambda^2 }}{m-\tf-\lambda}\\
    -\ii\\
     \frac{\ii\left(\kappa -\sqrt{(m-\tf)^2+\kappa^2 -\lambda^2 }\right)}{m-\tf-\lambda}\\ 1
\end{pmatrix}\,,\, 
w_{+} \approx \begin{pmatrix}
    \frac{-\kappa +\sqrt{(m+\tf)^2+\kappa^2 -\lambda^2 }}{m+\tf-\lambda}\\
    \ii\\
     \frac{\ii\left(\kappa -\sqrt{(m+\tf)^2+\kappa^2 -\lambda^2}\right)}{m+\tf-\lambda}\\ 1
\end{pmatrix}\,.
\label{Qeigenvec}\ee
We have neglected the terms which lead to exponentially small corrections in the variation of the $\eta$ function. For example, we can neglect $\mathcal{O}(\ta^2)$ terms and set $\tf = \tv$, neglecting the $\mathcal{O}(\tv - \tf)$ terms. 
All equations until the end of this subsection will be written up to such terms.

Edge states correspond to linear combinations of $w_-$ and $w_+$ belonging to
the kernel of the boundary projector $\Pi_{\varepsilon}$, which reads 
\begin{equation}
\operatorname{span}\left\{ z_{+},z_{-} \right\} = \operatorname{span}\left\{ \begin{pmatrix}
     -1 \\\ii\varepsilon \\ 0 \\ 0
\end{pmatrix}, \begin{pmatrix}
     0 \\ 0 \\  -1 \\\ii\varepsilon
\end{pmatrix} \right\}.
\end{equation}
Finding these linear combinations together with the restrictions on the parameters which ensure their existence reduces to studying the equation
\begin{equation}
\operatorname{det}(w_+, w_-,z_+, z_-) = 0\,.
\label{detEq}\end{equation} 
Whenever edge modes exist, their decay rate is governed by the smallest positive eigenvalue of $Q$, i.e., they decay as $e^{-\mu x^\mathbf{n}}$, where $\mu = \operatorname{min}\{ | m \pm \tv |  \}$ and are well-localized near the boundary as long as $\mu R$ is sufficiently large.

The results are as follows.  For 
$\varepsilon = -\operatorname{sgn}(m-\tv)$ and $|m|<\tv$, we find there is a single edge state with 
\begin{equation}
\lambda = -\varepsilon\left( \kappa   +\frac{1}{2}\ta \right)\,.\label{lambdaOaCase1}
\end{equation}
For the same condition on $\varepsilon$ and $|m|>\Tilde{v}$, we find instead  there are two edge states corresponding to
\begin{equation}
\lambda = -\varepsilon \left(\kappa + \left(\frac{1}{2}-\frac{\sqrt{m^2-\tv^2}}{2 m}\right)\ta \right)\quad \mbox{and}\quad
\lambda=-\varepsilon \left(\kappa + \left(\frac{1}{2}+\frac{\sqrt{m^2-\tv^2}}{2 m}\right)\ta \right).\label{lambdaOaCase2}
\end{equation}
For the opposite choice of the parameter $\varepsilon$ in the boundary projector,
$\varepsilon = +\operatorname{sgn}(m-\tv)$, and $|m|>\tv$ there are no edge states and for $|m| < \tv$ there is a single edge state with
\be
\lambda = -\varepsilon  \left(\kappa+\frac{1}{2}\ta \right)\,. \label{lambdaOa1Check}
\ee

The spectra obtained above allow us to express the boundary Hamiltonians as a Hamiltonian for a one-component field on $S_R^1$ with charge $\mathfrak{g}$ interacting with $A_\parallel^{[\alpha]}=A_\parallel+n/R$,
\begin{equation}
G(\mathfrak{g})=\ii (\partial_\parallel -\ii\mathfrak{g} (A_\parallel+n/R))\,.
\end{equation}
Thus, for $|m| < \tv$ the boundary Hamiltonian reads
\begin{equation}
H_{\mathrm{b}}= - \varepsilon\, G\left( \frac e2 \right)\,.
\end{equation}
For $\varepsilon = -\operatorname{sgn}(m-\tv)$ and $|m|>\tv$ one has
\begin{equation}
H_{\mathrm{b}}= - \varepsilon\,\left[  G\left(  e\left(\frac{1}{2}-\frac{\sqrt{m^2-\tv^2}}{2 m}\right)  \right)  \oplus  G\left(e \left(\frac{1}{2}+\frac{\sqrt{m^2-\tv^2}}{2 m}\right)  \right)\right].
\end{equation}
Using this identification, we compute the variations of $\eta$ functions with the help of (\ref{vareta}) and (\ref{a0}) with $d=1$,
\begin{equation}
\delta\eta(0,G)=-\frac{2}{\sqrt{\pi}}a_0(\delta G, G^2)=-\frac{1}{\pi}\int_{S^1_R}\dd x^\parallel\, \delta(\mathfrak{g}(A_\parallel+n/R)) \,.\label{deltaetaG}
\end{equation}
Note that changing the sign in front of $H$ leads to changing the sign of $\eta(0,H)$. If  $H$ is a direct sum of two  Hamiltonians, the $\eta$ functions add up.

\subsection{{The $\eta$ Invariant on the Plane}}\label{sec:together}

Following the method presented  in section \ref{se:method}, we compute the $\eta$ invariant on the plane by subtracting the edge state contribution from the result on the disk.

Let us start with the case $|m|>\tv$. For $\varepsilon=-\operatorname{sgn}(m-\tv)$, we obtain by (\ref{deleee}) and (\ref{delHDR})  that 
\begin{eqnarray}
&&\delta \eta(0,H)_R=\frac{\varepsilon e}{2\pi}\int_{S_R^1}dx^\parallel \delta A_\parallel \nonumber\\
&&\qquad -\frac{\varepsilon e}{\pi} \int_{S_R^1}dx^\parallel \delta \left[  
\left(\frac{1}{2}-\frac{\sqrt{m^2-\tv^2}}{2 m}\right)(A_\parallel+n/R) +  
\left(\frac{1}{2}+\frac{\sqrt{m^2-\tv^2}}{2 m}\right)(A_\parallel+n/R)\right] \nonumber\\
&&\qquad = -\frac{\varepsilon e}{2\pi}\int_{S_R^1}dx^\parallel \delta A_\parallel\,.
\label{case1}
\end{eqnarray}
Note that all terms containing variations of $m$ and $\tv$ cancel each other in the expression above.

For $\varepsilon=\operatorname{sgn}(m-\tv)$ there are no edge states. Thus the integral on the second line of (\ref{case1}) is absent and
\begin{equation}
\delta \eta(0,H)_R =\frac{\varepsilon e}{2\pi}\int_{S_R^1}dx^\parallel \delta A_\parallel \,.\label{case2}
\end{equation}
Equations (\ref{case1}) and (\ref{case2}) can be unified in a single expression which is valid for $|m|>\tv$ and does not depend on boundary conditions,
\begin{equation}
\delta \eta(0,H)_R =\operatorname{sgn}(m-\tv)\frac{ e}{2\pi}\int_{S_R^1}dx^\parallel \delta A_\parallel \,.\label{case12}
\end{equation}

By repeating the same steps for $|m|<\tv$ we obtain
\begin{equation}
\delta \eta(0,H)_R=0\,.\label{case3}
\end{equation}
The two cases considered above, (\ref{case12}) and (\ref{case3}) can be unified in a single relation,
\begin{equation}
\delta \eta(0,H)_R =\frac 12 [\operatorname{sgn}(m-\tv)+ \operatorname{sgn}(m+\tv)]\frac{ e}{2\pi}
\int_{S_R^1}dx^\parallel \delta A_\parallel \,.\label{case123}
\end{equation}
Since the sign functions have vanishing smooth variations, the variational equation (\ref{case123}) can be integrated, yielding the following expression for the $\eta$ function\footnote{\begin{minipage}[t]{\columnwidth}Note that the value of $|\phi|$ at infinity plays a role similar to the chemical potential, cf. Eq. (10.19) in \cite{Niemi:1984vz}, although it refers to a different system and involves the calculation of a different quantity.\end{minipage}}
\begin{equation}
    \eta(0,H) = \frac{\operatorname{sgn}(m+e \sqrt{2} v) +\operatorname{sgn}(m-e \sqrt{2} v)}{2}  \left( -\frac e{4\pi} \int_{\mathbb{R}^2} \dd^2x\, \epsilon^{jk}F_{jk} \right),
 \label{etaCandidate2} \end{equation}
where we took the limit $R\to\infty$. 

Finally, we use the expressions (\ref{Neta}) and (\ref{ANOn}) to identify the vacuum expectation value of the fermion number as
\begin{equation}
\mathcal{N}(H) =\frac{\operatorname{sgn}(m+e \sqrt{2} v) +\operatorname{sgn}(m-e \sqrt{2} v)}{4}\, n \,.\label{Nfinal}
\end{equation}
We stress that (\ref{Nfinal}) does not depend on  $\varepsilon$, i.e. is independent of the choice of boundary conditions. This provides a nontrivial  sanity check for the whole procedure.

One may notice that solutions to the variational equation (\ref{case123}) can involve an integration constant. Since our method breaks down when $m= \pm e \sqrt{2} v$, in fact three integration constants are needed, one for each of the intervals $m \in(-\infty,-e \sqrt{2} v),$ $m \in(-e \sqrt{2} v, e \sqrt{2} v)$, and $m \in(e \sqrt{2} v, \infty)$. However these three constants are fixed by considering $m \rightarrow \pm \infty$ (equivalently, $v \rightarrow 0)$ and $m=0$. These limits are considered in Sec. \ref{sec:dis} where the choices leading to (\ref{Nfinal}) are confirmed.

\section{Discussion}\label{sec:dis}
In the limit $v\to 0$, which means a vanishing scalar field, Eq.\ (\ref{Nfinal}) yields the expression
\begin{equation}
\mathcal{N}=\frac{n\, \operatorname{sgn}(m)}{2}\,,
\label{NS}
\end{equation}
obtained by Niemi and Semenoff \cite{Niemi:1983rq} for a planar fermion in an external magnetic field. This is an important consistency check of our calculations. (Note that a similar expression is valid for the ANO vortex with the couplings of Ref.\cite{Chamon:2007hx}.) Our result (\ref{Nfinal}) has a  richer structure. While both expressions have half-integer values for an odd topological number $n$,  in our model $\mathcal{N}$ also \emph{jumps} by a half-integer number when $\pm m$ crosses $e\sqrt{2}v$. 

Let us stress a peculiar property of the edge states. For $|m|<\tv$ the edge states have electric charge $\mathfrak{g}= e/2$, while for $|m|>\tv$ and $\varepsilon = -\operatorname{sgn}(m-\tv)$ the edge states have charges which change continuously from $0$ to $e/2$ and from $e/2$ to $e$, respectively.
Edge states with fractional charge play an important role in the theory of Fractional Quantum Hall Effect \cite{MacDonald:1990zz,Lopez:1998ih}.
The $e/2$ charged edge states appear in graphene nanoribbons \cite{Jeong_2019}.
%
% \old{Also, note the edge states which have fractional charges depending on the relations between $m$ and $v$ with potential applications in condensed matter physics \cite{MacDonald:1990zz,Lopez:1998ih,Jeong_2019}.
% \textbf{[Sylvain: More details here?]}}
%
Our results also have implications for the fermionic zero-mode structure of the ANO vortex on $\mathbb{R}^2$. For large $|m|$, the coupling to $\phi$ can be neglected. One can write $H^2(\phi=0)=H^2(\phi=0,m=0)+m^2>0$ which demonstrates that there are no zero modes for large $|m|$. The discontinuity of $\eta(0,H)$ then indicates that there are $n$ fermionic zero modes on $\mathbb{R}^2$ for $-e\sqrt{2}v< m<  e\sqrt{2}v$. 

For $m=0$, the presence of zero modes can be independently established as follows. Let us define
\begin{equation}
\Beta = \begin{pmatrix}
\beta & 0 \\ 0 & -\beta
\end{pmatrix}.
\end{equation}
It is easy to see that $\{ H(m=0),\Beta\}=0$. By using the usual index theory arguments, see Sec. 5.9 of Ref. \cite{Fursaev:2011zz}, we derive that the non-zero spectrum of $H$ is symmetric, thus yielding a vanishing $\eta$ function, in agreement with Eq. (\ref{etaCandidate2}). The zero modes can be separated in the eigenspaces of $\Beta$ with the eigenvalues $\pm 1$. Let $N_\pm$ be the dimension of the space of zero modes with $\Beta=\pm 1$. Then,
\begin{equation}
N_+-N_-\equiv \mbox{Ind}(H(m=0);\Beta)=a_2(\Beta,H^2(m=0))=\frac e{4\pi}\int \dd^2x \epsilon^{jk}F_{jk}=n\,.
\end{equation}
An expression for $a_2$ can be found in \cite{Vassilevich:2003xt}, for example. Thus, for $m=0$ the operator $H$ has at least $n$ zero modes on $\mathbb{R}^2$. The case of generic $m$ requires a different technique and goes beyond the scope of present work. We hope to address this problem in a future publication.

\subsection*{Acknowledgments}
This work was supported in parts by the S\~ao Paulo Research Foundation (FAPESP) through the grant 2021/10128-0. Besides, LS was supported by the grant 2023/11293-0 of FAPESP,  and DV was supported by the National Council for Scientific and Technological Development (CNPq), grant 304758/2022-1. 

% \appendix
% \section{Other boundary conditions}\label{sec:other}

\bibliographystyle{JHEP}
\bibliography{parity}

\providecommand{\href}[2]{#2}\begingroup\raggedright\begin{thebibliography}{10}

\bibitem{Fresneda:2023wub}
R.~Fresneda, L.~de~Souza, and D.~Vassilevich, {\it {Edge states and the \ensuremath{\eta} invariant}},  {\em Phys. Lett. B} {\bf 844} (2023) 138098, [\href{http://arxiv.org/abs/2305.13606}{{\tt arXiv:2305.13606}}].

\bibitem{Jackiw:1975fn}
R.~Jackiw and C.~Rebbi, {\it {Solitons with Fermion Number 1/2}},  {\em Phys. Rev. D} {\bf 13} (1976) 3398--3409.

\bibitem{Goldstone:1981kk}
J.~Goldstone and F.~Wilczek, {\it {Fractional Quantum Numbers on Solitons}},  {\em Phys. Rev. Lett.} {\bf 47} (1981) 986--989.

\bibitem{Jackiw:1981wc}
R.~Jackiw and J.~R. Schrieffer, {\it {Solitons with Fermion Number 1/2 in Condensed Matter and Relativistic Field Theories}},  {\em Nucl. Phys. B} {\bf 190} (1981) 253--265.

\bibitem{Paranjape:1983dy}
M.~B. Paranjape and G.~W. Semenoff, {\it {Spectral Asymmetry, Trace Identities and the Fractional Fermion Number of Magnetic Monopoles}},  {\em Phys. Lett. B} {\bf 132} (1983) 369--373.

\bibitem{Niemi:1984vz}
A.~J. Niemi and G.~W. Semenoff, {\it {Fermion Number Fractionization in Quantum Field Theory}},  {\em Phys. Rept.} {\bf 135} (1986) 99.

\bibitem{Niemi:1983rq}
A.~J. Niemi and G.~W. Semenoff, {\it {Axial Anomaly Induced Fermion Fractionization and Effective Gauge Theory Actions in Odd Dimensional Space-Times}},  {\em Phys. Rev. Lett.} {\bf 51} (1983) 2077.

\bibitem{Abrikosov:1956sx}
A.~A. Abrikosov, {\it {On the Magnetic properties of superconductors of the second group}},  {\em Sov. Phys. JETP} {\bf 5} (1957) 1174--1182.

\bibitem{Nielsen:1973cs}
H.~B. Nielsen and P.~Olesen, {\it {Vortex Line Models for Dual Strings}},  {\em Nucl. Phys. B} {\bf 61} (1973) 45--61.

\bibitem{Kim:2025ien}
Y.~Kim, S.~Jeon, and H.~Song, {\it {Charged Vortex in Superconductor}},  \href{http://arxiv.org/abs/2505.04359}{{\tt arXiv:2505.04359}}.

\bibitem{Vassilevich:2003xk}
D.~V. Vassilevich, {\it {Quantum corrections to the mass of the supersymmetric vortex}},  {\em Phys. Rev. D} {\bf 68} (2003) 045005, [\href{http://arxiv.org/abs/hep-th/0304267}{{\tt hep-th/0304267}}].

\bibitem{Rebhan:2003bu}
A.~Rebhan, P.~van Nieuwenhuizen, and R.~Wimmer, {\it {Nonvanishing quantum corrections to the mass and central charge of the N = 2 vortex and BPS saturation}},  {\em Nucl. Phys. B} {\bf 679} (2004) 382--394, [\href{http://arxiv.org/abs/hep-th/0307282}{{\tt hep-th/0307282}}].

\bibitem{Bordag:2003at}
M.~Bordag and I.~Drozdov, {\it {Fermionic vacuum energy from a Nielsen-Olesen vortex}},  {\em Phys. Rev. D} {\bf 68} (2003) 065026, [\href{http://arxiv.org/abs/hep-th/0305002}{{\tt hep-th/0305002}}].

\bibitem{AlonsoIzquierdo:2004ru}
A.~Alonso~Izquierdo, W.~Garcia~Fuertes, M.~de~la Torre~Mayado, and J.~Mateos~Guilarte, {\it {Quantum corrections to the mass of self-dual vortices}},  {\em Phys. Rev. D} {\bf 70} (2004) 061702, [\href{http://arxiv.org/abs/hep-th/0406129}{{\tt hep-th/0406129}}].

\bibitem{Graham:2004jb}
N.~Graham, V.~Khemani, M.~Quandt, O.~Schroeder, and H.~Weigel, {\it {Quantum QED flux tubes in 2+1 and 3+1 dimensions}},  {\em Nucl. Phys. B} {\bf 707} (2005) 233--277, [\href{http://arxiv.org/abs/hep-th/0410171}{{\tt hep-th/0410171}}].

\bibitem{Alonso-Izquierdo:2016bqf}
A.~Alonso-Izquierdo, J.~Mateos~Guilarte, and M.~de~la Torre~Mayado, {\it {Quantum magnetic flux lines, BPS vortex zero modes, and one-loop string tension shifts}},  {\em Phys. Rev. D} {\bf 94} (2016), no.~4 045008, [\href{http://arxiv.org/abs/1605.09175}{{\tt arXiv:1605.09175}}].

\bibitem{Graham:2022adn}
N.~Graham and H.~Weigel, {\it {Quantum energies of BPS vortices in D=2+1 and D=3+1}},  {\em Phys. Rev. D} {\bf 106} (2022), no.~7 076013, [\href{http://arxiv.org/abs/2207.04960}{{\tt arXiv:2207.04960}}].

\bibitem{Chamon:2007hx}
C.~Chamon, C.-Y. Hou, R.~Jackiw, C.~Mudry, S.-Y. Pi, and G.~Semenoff, {\it {Electron fractionalization for two-dimensional Dirac fermions}},  {\em Phys. Rev. B} {\bf 77} (2008) 235431, [\href{http://arxiv.org/abs/0712.2439}{{\tt arXiv:0712.2439}}].

\bibitem{Jackiw:1981ee}
R.~Jackiw and P.~Rossi, {\it {Zero Modes of the Vortex - Fermion System}},  {\em Nucl. Phys. B} {\bf 190} (1981) 681--691.

\bibitem{Almeida:2021lks}
C.~Almeida, A.~Alonso-Izquierdo, R.~Fresneda, J.~Mateos~Guilarte, and D.~Vassilevich, {\it {Nontopological fractional fermion number in the Jackiw-Rossi model}},  {\em Phys. Rev. D} {\bf 103} (2021), no.~12 125015, [\href{http://arxiv.org/abs/2103.06826}{{\tt arXiv:2103.06826}}].

\bibitem{Alonso-Izquierdo:2019tms}
A.~Alonso-Izquierdo, R.~Fresneda, J.~Mateos~Guilarte, and D.~Vassilevich, {\it {Soliton Fermionic number from the heat kernel expansion}},  {\em Eur. Phys. J. C} {\bf 79} (2019), no.~6 525, [\href{http://arxiv.org/abs/1905.09030}{{\tt arXiv:1905.09030}}].

\bibitem{MacDonald:1990zz}
A.~H. MacDonald, {\it {Edge states in the fractional-quantum-Hall-effect regime}},  {\em Phys. Rev. Lett.} {\bf 64} (1990) 220--223.

\bibitem{Lopez:1998ih}
A.~Lopez and E.~H. Fradkin, {\it {Universal structure of the edge states of the fractional quantum Hall states}},  {\em Phys. Rev. B} {\bf 59} (1999) 15323, [\href{http://arxiv.org/abs/cond-mat/9810168}{{\tt cond-mat/9810168}}].

\bibitem{AlvarezGaume:1984nf}
L.~Alvarez-Gaume, S.~Della~Pietra, and G.~W. Moore, {\it {Anomalies and Odd Dimensions}},  {\em Annals Phys.} {\bf 163} (1985) 288.

\bibitem{Atiyah:1980jh}
M.~F. Atiyah, V.~K. Patodi, and I.~M. Singer, {\it {Spectral asymmetry and Riemannian geometry. III}},  {\em Math. Proc. Cambridge Phil. Soc.} {\bf 79} (1976) 71--99.

\bibitem{Gilkey:1984}
P.~B. Gilkey, {\em {Invariance theory, the heat equation, and the Atiyah-Singer index theorem}}.
\newblock Publish or Perish, Wilmington, 1984.

\bibitem{Kurkov:2018pjw}
M.~Kurkov and D.~Vassilevich, {\it {Gravitational parity anomaly with and without boundaries}},  {\em JHEP} {\bf 03} (2018) 072, [\href{http://arxiv.org/abs/1801.02049}{{\tt arXiv:1801.02049}}].

\bibitem{Vassilevich:2003xt}
D.~V. Vassilevich, {\it {Heat kernel expansion: User's manual}},  {\em Phys. Rept.} {\bf 388} (2003) 279--360, [\href{http://arxiv.org/abs/hep-th/0306138}{{\tt hep-th/0306138}}].

\bibitem{Jeong_2019}
Y.~H. Jeong, S.-R. Eric~Yang, and M.-C. Cha, {\it Soliton fractional charge of disordered graphene nanoribbon},  {\em Journal of Physics: Condensed Matter} {\bf 31} (2019), no.~26 265601.

\bibitem{Fursaev:2011zz}
D.~Fursaev and D.~Vassilevich, {\em {Operators, Geometry and Quanta}}.
\newblock Theoretical and Mathematical Physics. Springer, Berlin, Germany, 2011.

\end{thebibliography}\endgroup

\end{document}